\documentclass[iop,apjl]{emulateapj}
\usepackage{CJK}
\usepackage{natbib}
\usepackage{graphicx}
\usepackage{apjfonts}

\shorttitle{Water vapor in nearby infrared galaxies}
\shortauthors{Yang et al.}

\begin{document}
\begin{CJK*}{UTF8}{gbsn}

\title{Water vapor in nearby infrared galaxies as probed by {\it Herschel}$^\star$}

\author{
 Chentao Yang (杨辰涛) \altaffilmark{1,2,3}, 
 Yu Gao (高煜) \altaffilmark{2}, 
 A. Omont \altaffilmark{4,5},
 Daizhong Liu (刘岱钟) \altaffilmark{2,3}, 
 K. G. Isaak \altaffilmark{6},
 D. Downes \altaffilmark{7},
 P. P. van der Werf \altaffilmark{8},
 and Nanyao Lu \altaffilmark{9}
  }

\altaffiltext{1}{Department of Astronomy, Beijing Normal University, Beijing 100875, China} 
\altaffiltext{2}{Purple Mountain Observatory/Key Lab of Radio Astronomy, Chinese Academy of Sciences, Nanjing 210008, China}
\altaffiltext{3}{University of Chinese Academy of Sciences, Beijing, China}
\altaffiltext{4}{Institut d'Astrophysique de Paris, UPMC Universit\'{e} Paris 06, UMR7095, F-75014, Paris, France}
\altaffiltext{5}{CNRS, UMR7095, Institut d'Astrophysique de Paris, F-75014, Paris, France}
\altaffiltext{6}{ESA Astrophysics Missions Division, ESTEC, PO Box 299, 2200 AG Noordwijk, The Netherlands}
\altaffiltext{7}{Institut de Radioastronomie Millim{\'e}trique (IRAM), 300 rue de la Piscine, F-38406 Saint-Martin d'H{\`e}res, France}
\altaffiltext{8}{Leiden Observatory, Leiden University, Post Office Box 9513, NL-2300 RA Leiden, The Netherlands}
\altaffiltext{9}{Infrared Processing and Analysis Center, California Institute of Technology, MS 100-22, Pasadena, CA 91125, USA}
\altaffiltext{$\star$}{{\it Herschel} is an ESA space observatory with science instruments provided by European-led Principal Investigator consortia and with important participation from NASA.}

\begin{abstract}
We report the first systematic study of the submillimeter water vapor rotational
emission lines in infrared (IR) galaxies based on the Fourier Transform 
Spectrometer (FTS) data of {\it Herschel} SPIRE. 
Among the 176 galaxies with publicly available FTS data,
45 have at least one H$_2$O emission line detected. 
The H$_2$O line luminosities range from $\sim 1 \times 10^5$ $L_{\odot}$ to 
$\sim 5 \times 10^7 L_{\odot}$ while the total IR luminosities ($L_\mathrm{IR}$)
have a similar spread ($\sim 1-300 \times 10^{10} L_{\odot}$).
In addition, emission lines of H$_2$O$^+$ and H$_2^{18}$O are also detected.
H$_2$O is found, for most galaxies, to be the strongest molecular emitter 
after CO in FTS spectra.  
The luminosity of the five most important H$_2$O lines is 
near-linearly correlated with $L_\mathrm{IR}$, no matter whether
strong active galactic nucleus signature is present or not. 
However, the luminosity of H$_2$O($2_{11}-2_{02}$) and H$_2$O($2_{20}-2_{11}$) 
appears to increase slightly faster than linear with $L_\mathrm{IR}$.
Although the slope turns out to be slightly steeper when $z\sim 2-4$ ULIRGs 
are included, the correlation is still closely linear.
We find that $L_\mathrm{H_2O}$/$L_\mathrm{IR}$ decreases with increasing 
$f_{25}$/$f_{60}$, but see no dependence on $f_{60}$/$f_{100}$, possibly 
indicating that very warm dust contributes little to 
the excitation of the submillimeter H$_2$O lines. 
The average spectral line energy distribution (SLED) of the entire sample is 
consistent with individual SLEDs and the IR pumping plus collisional 
excitation model, showing that the strongest lines are 
H$_2$O($2_{02}-1_{11}$) and H$_2$O($3_{21}-3_{12}$).
\end{abstract}


\section{Introduction}
H$_2$O can be one of the most abundant oxygen molecular carriers besides CO in the warm 
interstellar gas (but it is mostly locked in icy interstellar dust grains in cold regions
of the Galaxy, e.g., \citealp{Melnick2005, vand2011}).
Nevertheless, the study of the rotational H$_2$O line is always far more challenging 
than CO at low redshift.
The main difficulty is from the contamination of the H$_2$O in the Earth atmosphere.
However, some pioneering research with the \textit{Infrared Space Observatory} 
(covering $\sim2-200\,\mu$m) \citep{Kessler1996} of both star-forming regions within our Galaxy, 
like Orion \citep{Harwit1998}, and nearby galaxies, like Arp220 \citep{gonz2004},
NGC253 and NGC1068 \citep{Goi2005}, and Mrk231 \citep{gonz2008}, revealed that
H$_2$O lines likely trace the local infrared radiation field (IRF) directly, 
thus provide a unique diagnostic probing the physical and chemical processes unlike 
other gas tracers like CO.
The {\it Herschel Space Observatory} \citep{Pilbratt2010} with great improvement of sensitivity, 
angular resolution and band coverage offers an unprecedented opportunity
to study the submillimeter regime of galaxies without atmospheric contamination, 
thus provides unique chances to observe the H$_2$O lines
within the SPIRE band \citep[$194-672\,\mu$m,][]{Griffin2010}.

{\it Herschel} has revealed a wealth of submillimeter H$_2$O lines in, 
e.g., Mrk231 \citep[][G-A10 hereafter]{vandw2010, gonz2010}, 
Arp220 \citep{rang2011, gonz2012, gonz2013},
NGC4418 \citep{gonz2012}, NGC1068 \citep{Spinoglio2012}, NGC6240 \citep{Meijerink2013} and 
M82 \citep{Kamenetzky2012}, from energy level $E_\mathrm{up}$/$k = 53$ K up to
$E_\mathrm{up}$/$k = 642$ K. 
Moreover, some detections from ground in high-$z$ ultra-luminous IR galaxies (ULIRGs) 
were also reported \citep[e.g.][]{omont2011, vandw2011, Combes2012, omont2013, Riechers2013}.
H$_2$O line strength is found to be comparable with neighboring high-$J$ CO
lines ($J = 8-7$ to $J = 13-12$) in these studies. 

By modeling the H$_2$O excitation and dust continuum in Mrk231, 
G-A10 interpreted that collisional excitation from a cool extended 
region (610 pc, 41 K) is responsible for part of the low-lying line excitation,
while IR pumping through far-IR photons by compact warm dense gas (120 pc, 95 K)
excites high-lying lines and part of low-lying lines.
The high abundance of H$_2$O can be explained as a consequence of 
shocks/cosmic rays and X-ray dominated regions (XDR) chemistry \citep{Meijerink2005}, and/or
an “undepleted chemistry” (G-A10).
Therefore, H$_2$O excitation is naturally linked to the local IRF, 
probing, e.g., the size and strength of the IR power source; 
tracing a different regime of gas than that of CO.
Hence it is important to have a systematic study of the H$_2$O lines in galaxies, 
for better understanding the gas excitation and physical processes within.

\section{The Sample and Data Reduction}
We used the {\it Herschel} Science Archive (HSA),
containing both the SPIRE//Fourier Transform Spectrometer \citep[FTS;][]{Naylor2010}
spectra at 450--1550GHz, and the PACS \citep{Poglitsch2010} images at 70, 100 and $160\,\mu$m. 
Our sample consists of 45 sources with at least one rotational H$_2$O 
transition detected among 176 nearby galaxies available.
The data are from 10 projects including {\it HerCULES} (PI: P. van der Werf) 
with an H$_2$O detection rate $\sim80$\% and {\it GOALS} (PI: N. Lu., 
a full list can be found here: 
{\small \texttt{http://sfig.pmo.ac.cn/\~{}yangcht/h2oSample.txt}}).
The typical SPIRE/FTS integration time is about several hours.

The data were reduced with HIPE v9 \citep{Ott2010}. 
Basic steps of spectral data reduction contain background removal using off-axis 
detector subtraction and flux calibration with Neptune and Uranus, when available. 
Deglitch, flat field, calibration through HIPE, and brightness drift subtraction 
with \texttt{Scanamorphos} \citep{Roussel2012} have been used to reduce PACS images.
All the H$_2$O emission line detections are above the $3\sigma$ level.
The instrumental sinc function has been adopted for the line 
fit using customized HIPE scripts, since the intrinsic line width 
is smaller than the instrumental resolution in most cases.
However, the flux could be underestimated by $\sim$20\% for few
sources with very broad linewidth like Arp220, 
it is still insignificant when we consider the line fitting error ($\sim20\%$), 
the main source of the errors.
Then we use the formula in \cite{solomon1992} to convert line
intensity ($I_\mathrm{H_2O}$) to $L_\mathrm{H_2O}$, taking the
luminosity distance $D_L$ in \cite{sanders2003} 
($H_0 = 75\,\mathrm{km\,s^{-1}\,Mpc^{-1}}$, 
$\Omega_{M} = 0.3$, and $\Omega_{\Lambda} = 0.7$; \citealp{Mould2000}).

After convolving \textit{Spitzer}/MIPS $24\,\mu$m, PACS 70, 100, and $160\,\mu$m images to 
match with the SPIRE beams \citep{Swinyard2010} following \cite{Aniano2011}, 
we determine whether the source is extended or not based on its radial profile as compared 
with that of the corresponding Gaussian point-spread functions (PSFs).
We use the total IR luminosities (8--1000 $\mu$m) from \cite{sanders2003} as 
the $L_\mathrm{IR}$ for point sources. 
For extended galaxies, in-beam $L_\mathrm{IR}$ is calculated
to ensure that $L_\mathrm{IR}$ and $L_\mathrm{H_2O}$ are spatially matched.
First, we take the weighting coefficients of \cite{Galametz2013} 
to combine MIPS 24, PACS 70, 100 and 160 $\mu$m images into composite maps. 
Then the in-beam flux ratio between in-aperture and that of the entire source
is derived with aperture photometries (FWHM of the Gaussian PSFs). 
It should be noticed that the practically measured area by SPIRE/FTS 
is not limited in the FWHM beam, we need an additional correction 
factor to account for this (D. Liu et al., in preparation).
Applying this factor, we can then obtain the corrected in-beam 
fraction of the $L_\mathrm{IR}$ for extended sources. 
$L_\mathrm{IR}$ matched with the SPIRE beam can thus be obtained
by applying this factor and the in-beam fraction
to the global $L_\mathrm{IR}$ in \cite{sanders2003}.
The full dataset containing $L_\mathrm{IR}$ and $L_\mathrm{H_2O}$
will be described in D. Liu et al. (in preparation).
Since we take the global flux density ratios of 25 -- $60\,\mu$m 
($f_{25}$/$f_{60}$) and 60 -- 100 $\mu$m ($f_{60}$/$f_{100}$) as
the IR colors \citep[flux densities from][]{sanders2003} in later 
analysis, we exclude the extended sources in these cases in order to keep 
the IR colors free from any contamination in spatial variations.

H$_2$O lines are also detected in the mapping mode data of M82, NGC1068 and NGC253. 
However, we have dropped M82 mapping mode data because the very weak detection is only 
in the central detector and we already have a robust detection in the single pointing mode. 
For NGC253, we add all the spectral data to obtain a global spectrum. 
For NGC1068, we have at least one H$_2$O line detection in seven detectors.
These sources are obviously extended and excluded in the IR color analysis.

 \begin{figure*}
 \epsscale{0.79}
 \plotone{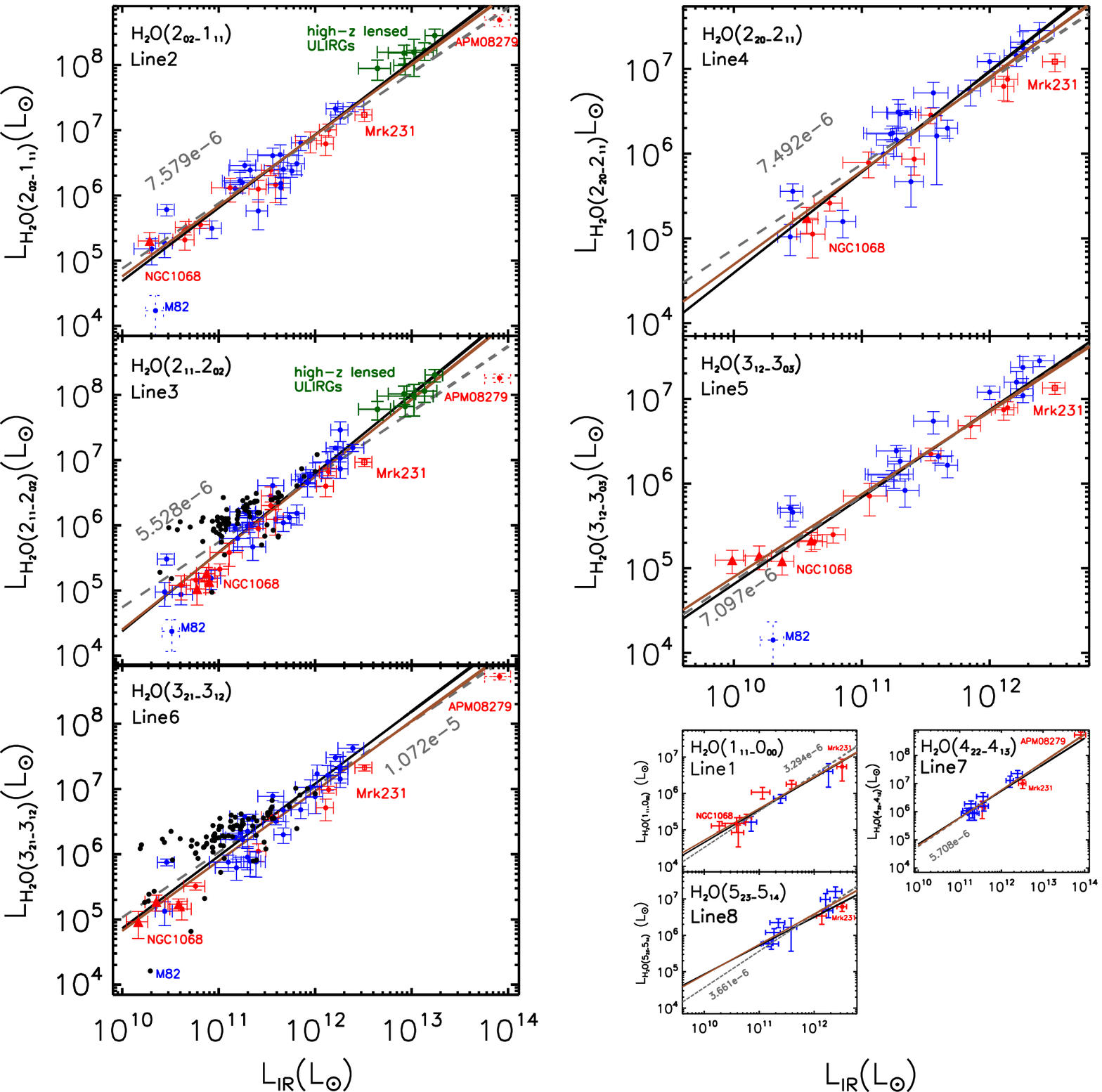}
 \caption{Correlation between $L_\mathrm{H_2O}$ and the corresponding $L_\mathrm{IR}$ of our sample.
 Fitted lines by \texttt{MPFIT} and \texttt{LINMIX\_ERR} are shown in black and brown lines, respectively,
 while gray lines are the linear fitting with fixed slope ($\alpha=1$).
 Red, blue, green and black dots represent strong-AGN, H{\footnotesize{II}}+mild-AGN-dominated
 galaxies, high-$z$ ULIRGs, and upper limits for non-detections, respectively. 
 Solid triangles are the mapping mode data of NGC1068. 
 Mrk231 is marked in red squares.
 M82 and APM08279+5255, marked with dashed error bars, are excluded from the fitting.
 (A color version of this figure is available in the online journal.)
 }
 \label{fit}
 \end{figure*}

\begin{table*}
\caption{Fitted Parameters of the Correlations between H$_2$O Lines and $L_\mathrm{IR}$ (Equation (\ref{eq1}))} 
\begin{center}
\begin{tabular}{lcccccc}
\hline 
\hline 
H$_{2}$O Line & $\nu_\mathrm{rest}$(GHz) & $\alpha_{\chi^{2}}$ & $\alpha_{\mathrm{Bayes}}$ & $\beta_{\chi^{2}}$ & $\beta_{\mathrm{Bayes}}$& {\footnotesize $\left\langle {L_\mathrm{H_2O}} /\ {L_\mathrm{IR}} \right\rangle $}\tabularnewline
\hline 
1, $1_{11}-0{_{00}}^a$ & 1113.343 & $\ensuremath{0.89\pm0.09}$ & $\ensuremath{\ensuremath{0.86\pm0.17}}$ & $\ensuremath{\ensuremath{-4.24\pm1.06}}$ & $\ensuremath{\ensuremath{-3.76\pm1.94}}$ & $\ensuremath{\ensuremath{3.29\times 10^{-6}}}$ \tabularnewline
2, $2_{02}-1_{11}$     &  987.927 & $\ensuremath{1.12\pm0.04}$ & $\ensuremath{\ensuremath{1.08\pm0.05}}$ & $\ensuremath{-6.52\pm0.47}$ & $\ensuremath{-6.07\pm0.59}$ & $\ensuremath{\ensuremath{7.58\times 10^{-6}}}$ \tabularnewline
3, $2_{11}-2_{02}$     &  752.033 & $\ensuremath{\ensuremath{1.21\pm0.04}}$ & $\ensuremath{\ensuremath{1.18\pm0.06}}$ & $\ensuremath{\ensuremath{-7.72\pm0.49}}$ & $\ensuremath{\ensuremath{-7.34\pm0.67}}$ & $\ensuremath{\ensuremath{5.53\times 10^{-6}}}$ \tabularnewline
4, $2_{20}-2_{11}$     & 1228.789 & $\ensuremath{\ensuremath{1.19\pm0.06}}$ & $\ensuremath{1.10\pm0.08}$ & $\ensuremath{\ensuremath{-7.30\pm0.69}}$ & $\ensuremath{\ensuremath{-6.33\pm0.97}}$ & $\ensuremath{\ensuremath{7.49\times 10^{-6}}}$ \tabularnewline
5, $3_{12}-3_{03}$     & 1097.365 & $\ensuremath{\ensuremath{1.03\pm0.04}}$ & $\ensuremath{0.98\pm0.06}$ & $\ensuremath{\ensuremath{-5.45\pm0.51}}$ & $\ensuremath{\ensuremath{-4.88\pm0.65}}$ & $\ensuremath{\ensuremath{7.10\times 10^{-6}}}$ \tabularnewline
6, $3_{21}-3_{12}$     & 1162.912 & $\ensuremath{\ensuremath{1.11\pm0.05}}$ & $\ensuremath{1.07\pm0.09}$ & $\ensuremath{-6.22\pm0.57}$ & $\ensuremath{-5.88\pm1.05}$ & $\ensuremath{\ensuremath{1.07\times 10^{-5}}}$ \tabularnewline
7, $4_{22}-4{_{13}}^a$ & 1207.639 & $\ensuremath{\ensuremath{0.94\pm0.12}}$ & $\ensuremath{\ensuremath{0.84\pm0.22}}$ & $\ensuremath{-2.91\pm1.12}$ & $\ensuremath{-3.43\pm2.51}$ & $\ensuremath{\ensuremath{5.71\times 10^{-6}}}$ \tabularnewline
8, $5_{23}-5{_{12}}^a$ & 1410.618 & $\ensuremath{\ensuremath{0.78\pm0.10}}$ & $\ensuremath{0.99\pm0.19}$ & $\ensuremath{\ensuremath{-4.56\pm1.37}}$ & $\ensuremath{\ensuremath{-4.93\pm2.30}}$ & $\ensuremath{\ensuremath{3.66\times 10^{-6}}}$ \tabularnewline
\hline 
\end{tabular}

\begin{tabular}{p{11.7cm}}
{\textbf{Notes.} $^{a}$: The resulting parameters of $1_{11}-0_{00}$, $4_{22}-4_{13}$ and $5_{23}-5_{12}$ 
contain large uncertainties due to the small sample size.
$\alpha_{\chi^{2}}$ and $\beta_{\chi^{2}}$ are the slope and intercept from $\chi^2$ fitting, 
while $\alpha_{\mathrm{Bayes}}$ and $\beta_{\mathrm{Bayes}}$ are from the Bayesian method.}
\end{tabular}

\end{center}
\label{line_fitting}
\end{table*}

\section{Results and Discussion}
In our sample we find that H$_2$O is the strongest molecular emitter after
high-$J$ CO ($J = 8-7$ to $J = 13-12$) in the SPIRE band.
In some cases ($\sim$13\%), e.g., ESO320-G030, the strength of 
H$_2$O($3_{21}-3_{12}$) is even stronger. 
Besides the H$_2$O emission lines, 
H$_2$O($1_{11}-0_{00}$) is detected in absorption in three sources, 
including Arp220 as reported by \cite{rang2011}.
H$_2$O$^{+}$ absorption lines were also detected in a few sources (D. Liu et al. in preparation).
In addition, emission lines of H$_2$O$^+$ and H$_2^{18}$O are detected (Section 3.3). 
Those ionic molecules are the intermediate species for the main route of gas-phase H$_2$O formation.

\subsection{Relation between $\mathrm{H_2O}$ and IR Luminosities}
The correlation between $L_\mathrm{H_2O}$ for different transitions 
and $L_{\mathrm{IR}}$ was analyzed by two different methods: 
a Bayesian approach, \texttt{LINMIX\_ERR} \citep{Kelly2007}, and 
the nonLinear $\chi^2$ fitting routine, \texttt{MPFIT} \citep{Markwardt2009}. 
In Figure \ref{fit} we plot the luminosities of 
H$_2$O($1_{11}-0_{00}$), H$_2$O($2_{02}-1_{11}$), 
H$_2$O($2_{11}-2_{02}$), H$_2$O($2_{20}-2_{11}$), 
H$_2$O($3_{12}-3_{03}$), H$_2$O($3_{21}-3_{12}$), 
H$_2$O($4_{22}-4_{13}$) and H$_2$O($5_{23}-5_{14}$) (lines 1 to 8 hereafter)
against the corresponding $L_{\mathrm{IR}}$.
In addition to our sample, we also include five high-$z$ ULIRGs
\cite[][see online Table 4]{omont2013} and HLSJ0918+5144 \citep{Combes2012} 
in our fit for line 2 and 3 (Figure \ref{fit}).
The QSO APM08279+5255 at $z = 3.9$ \citep{vandw2011} is also added for comparison.

The two fitting methods yield similar results in log--log space 
over four orders of magnitude of the luminosity range. 
The fit can be described as  
\begin{equation}
\label{eq1}
\log{L_\mathrm{H_2O}}=\alpha{\log{L_{\mathrm{IR}}}+\beta}.
\end{equation}
The derived parameters are listed in Table \ref{line_fitting}. 
All values of $\alpha$ are close to 1, i.e., a linear relation,
though the $\alpha$ given by the Bayesian approach are closer to linear.
However, the $\alpha$ of lines 3 and 4 are a bit higher than that of other lines.
This is weakly significant when we condsider the errors.
The $\alpha$ of lines 2 and 3 are consistent with \cite{omont2013}. 
As the slopes are close to linear, we perform an additional linear fit 
by fixing $\alpha=1$, and use $\chi^2$ fitting to determine the 
constant ratios between $L_\mathrm{H_2O}$ and $L_{\mathrm{IR}}$.
These ratios vary from $3.3{\times}10^{-6}$ for H$_2$O($1_{11}-0_{00}$) to $1.1{\times}10^{-5}$ 
for H$_2$O($3_{21}-3_{12}$) (see the gray dashed lines and text in
Figure \ref{fit} and Table \ref{line_fitting}). 
Because the detections of lines 1, 7 and 8 are not statistically significant, 
more data are needed to solidify the fits. 
In Figure \ref{fit}, we find most of the H$_2$O($2_{11}-2_{02}$) and 
H$_2$O($3_{21}-3_{12}$) upper limits for the non-detections are consistent with the correlation.
All the (U)LIRGs have a strong H$_2$O emission compatible with the correlation pointing
out to a rather large H$_2$O abundance as known in shocked regions \citep[e.g. G-A10;][]{Harwit1998}.
Unlike the case in the Orion Bar, the proto-typical photo-dissociation region (PDR),
where CO lines are a factor $\gtrsim50$ stronger than the H$_2$O lines, 
the high $\mathrm{H_2O/CO}$ ratio of most sources in our sample makes it unlikely 
that those strong H$_2$O emission originate in classical PDRs (e.g., G-A10).
The high $\mathrm{CO/H_2O}$ ratio in M82 \citep[$\sim40$;][]{weiss2010} indicates 
that it is dominated by classical PDRs, and thus has much weaker H$_2$O lines. 
As in \cite{weiss2010}, the H$_2$O lines in M82 are found 
to be very weak, nearly an order of magnitude below the correlation. 
It would be important to analyze the weak H$_2$O emission in other galaxies
like M82, but it is outside the scope of this work.
Therefore, we excluded M82 from our fit. 
Additionally, when we fit the correlations for $L_{\mathrm{H_2O}-2}$ and $L_{\mathrm{H_2O}-3}$ 
without high-$z$ ULIRGs, we get slightly lower slopes. 
This means that high-$z$ ULIRGs at the high $L_{\mathrm{IR}}$ end have slightly higher 
$L_{\mathrm{H_2O}}$/$L_{\mathrm{IR}}$.

The linear correlation could be the result of the very intense far-IR radiation via IR pumping. 
After the absorption of far-IR photons, 
the upper level H$_2$O molecules cascade toward the lines we observed in an 
approximately constant fraction.
Thus the H$_2$O luminosity should be linearly correlated with the IR emission. 
Though detailed excitation modeling is needed, this linear correlation already 
shows the importance of IR pumping.

\begin{table*}
\caption{Correlations between $(L_\mathrm{H_2O}$/$L_\mathrm{IR})$/$L_\mathrm{IR}$, 
         Different H$_2$O Line Ratios and IR Colors, $L_\mathrm{IR}$.} 
\begin{center}
\begin{tabular}{lccccc}
\hline 
\hline 
{  Ratio} & {  $R_{\log(f_{25}/f_{60})}$} & {  $R_{\log(f_{60}/f_{100})}$} & {  $R_{\log(L_{\mathrm{IR}})}$} & {  $\left\langle \mathrm{strong-AGN}\right\rangle $} & {  $\left\langle\mathrm{H\mbox{\scriptsize{II}}+mild-AGN}\right\rangle$}\tabularnewline
\hline 
{ 2/IR, $2_{02}-1_{11}$/IR}       & ${-0.51}$ &    ${-0.25}$ & ${-0.10}$ & {6.4$\times$10$^{-6}$} & {9.2$\times$10$^{-6}$}\tabularnewline
{ 3/IR, $2_{11}-2_{02}$/IR}       & ${-0.51}$ & \phs${0.17}$ & ${-0.39}$ & {3.8$\times$10$^{-6}$} & {5.8$\times$10$^{-6}$}\tabularnewline
{ 4/IR, $2_{20}-2_{11}$/IR}       & ${-0.55}$ &    ${-0.03}$ & ${-0.21}$ & {5.0$\times$10$^{-6}$} & {9.2$\times$10$^{-6}$}\tabularnewline
{ 5/IR, $3_{12}-3_{03}$/IR}       & ${-0.40}$ &    ${-0.02}$ & ${-0.18}$ & {5.1$\times$10$^{-6}$} & {9.4$\times$10$^{-6}$}\tabularnewline
{ 6/IR, $3_{21}-3_{12}$/IR}       & ${-0.67}$ &    ${-0.07}$ & ${-0.17}$ & {6.7$\times$10$^{-6}$} & {10.8$\times$10$^{-6}$}\tabularnewline
{ 2/3, $2_{02}-1_{11}$/$2_{11}-1_{02}$}  & \phs${0.16}$ &    ${-0.50}$ &    ${-0.38}$ & ${1.50}$ & ${1.52}$\tabularnewline
{ 2/4, $2_{02}-1_{11}$/$2_{20}-2_{11}$}  & \phs${0.16}$ &    ${-0.38}$ &    ${-0.47}$ & ${1.09}$ & ${1.09}$\tabularnewline
{ 2/5, $2_{02}-1_{11}$/$3_{12}-3_{03}$}  &    ${-0.11}$ &    ${-0.28}$ & \phs${0.25}$ & ${1.14}$ & ${0.88}$\tabularnewline
{ 2/6, $2_{02}-1_{11}$/$3_{21}-3_{12}$}  & \phs${0.56}$ &    ${-0.07}$ &    ${-0.20}$ & ${0.87}$ & ${0.72}$\tabularnewline
{ 3/4, $2_{11}-2_{02}$/$2_{20}-2_{11}$}  & \phs${0.10}$ & \phs${0.35}$ & \phs${0.08}$ & ${0.61}$ & ${0.63}$\tabularnewline
{ 3/5, $2_{11}-2_{02}$/$3_{12}-3_{03}$}  &    ${-0.26}$ & \phs${0.14}$ & \phs${0.47}$ & ${0.62}$ & ${0.52}$\tabularnewline
{ 3/6, $2_{11}-2_{02}$/$3_{21}-3_{12}$}  & \phs${0.47}$ & \phs${0.11}$ & \phs${0.15}$ & ${0.50}$ & ${0.48}$\tabularnewline
{ 4/5, $2_{20}-2_{11}$/$3_{12}-3_{03}$}  &    ${-0.15}$ &    ${-0.11}$ & \phs${0.38}$ & ${1.00}$ & ${0.58}$\tabularnewline
{ 4/6, $2_{20}-2_{11}$/$3_{21}-3_{12}$}  & \phs${0.37}$ & \phs${0.09}$ & \phs${0.19}$ & ${0.73}$ & ${0.62}$\tabularnewline
{ 5/6, $3_{12}-3_{03}$/$3_{21}-3_{12}$}  & \phs${0.53}$ & \phs${0.17}$ &    ${-0.21}$ & ${0.69}$ & ${0.65}$\tabularnewline
\hline 
\end{tabular}

\begin{tabular}{p{12cm}}
{\textbf{Notes.} $R_{\log(f_{25}/f_{60})}$, $R_{\log(f_{60}/f_{100})}$
and $R_{(L_{\mathrm{IR}})}$ are the correlation coefficients between 
$\log[(L_\mathrm{H_2O}$/$L_\mathrm{IR})$/$L_\mathrm{IR}$],  
$\log[(L_\mathrm{H_2O}$/$L_\mathrm{IR})_{a}$/$(L_\mathrm{H_2O}$/$L_\mathrm{IR})_{b}]$ 
and $\log(f_{25}/f_{60})$, 
$\log(f_{60}/f_{100})$, $\log(L_{\mathrm{IR}})$ respectively (see text). 
$\left\langle \mathrm{strong-AGN}\right\rangle$ and $\left\langle \mathrm{H\mbox{\scriptsize{II}}+mild-AGN}\right\rangle$ are the error 
weighted mean values of 
$(L_\mathrm{H_2O}$/$L_\mathrm{IR})$/$L_\mathrm{IR}$ and
$(L_\mathrm{H_2O}$/$L_\mathrm{IR})_{a}$/$(L_\mathrm{H_2O}$/$L_\mathrm{IR})_{b}$
for strong-AGN and star-forming dominated galaxies (see text), respectively.} 
\end{tabular}

\end{center}
\label{line_ratio}
\end{table*}

Using the NASA/IPAC Extragalactic Database (NED), we have separated our sample
into two groups: optically identified strong-active galactic nucleus (AGN)
dominated (Seyfert types 1 and 2) and
star-forming-dominated galaxies possibly with mild AGNs 
(classes H{\footnotesize{II}}, 
composite and LINER of \cite{Kewley2006}, 
hereafter "H{\footnotesize{II}}+mild-AGN"), 
as red and blue points in Figure \ref{fit}, respectively. 
There is no obvious difference between these two groups and they both 
exhibit similar correlations.
This implies that both strong-AGN and H{\footnotesize{II}}+mild-AGN
sources behave similarly in H$_2$O emission, 
and a strong AGN may have little impact on the H$_2$O excitation.
Although the number of statistics is small, 
the detection rate of H{\footnotesize{II}}+mild-AGN ($\sim3.2$\%) is lower 
than strong AGN ($\sim12.4$\%) for H$_2$O($1_{11}-0_{00}$).
The remaining H$_2$O lines have comparable detection rates of both kinds,
and line 2 and 3 have the highest detection rate of about 30\%.
The absence of and apparent significant AGN contribution
indicates that an AGN may not be the main power source of the H$_2$O excitation.
The origin of such abundant H$_2$O reservoir might thus
favor an undepleted chemistry or shocks/cosmic rays rather than XDR chemistry (G-A10).

\begin{figure}
 \epsscale{1.02}
 \plotone{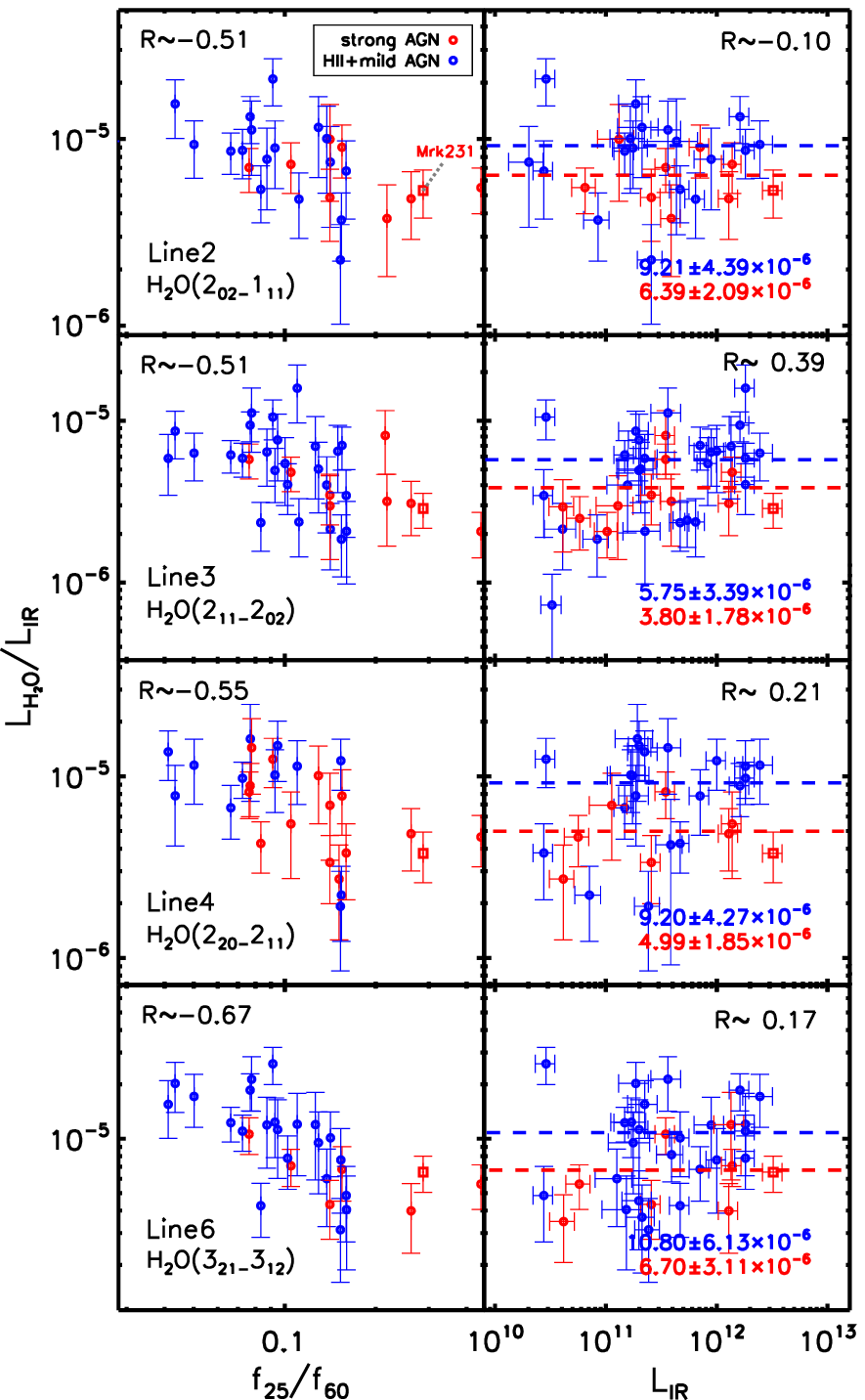}
 \caption{$L_{\mathrm{H}_2\mathrm{O}}$/$L_{\mathrm{IR}}$ versus
 $f_{25}$/$f_{60}$ and $L_{\mathrm{IR}}$, respectively. 
 From top to bottom, each row displays the values of lines 2,3,4 \& 6 as examples.
 Averaged values of $L_{\mathrm{H}_2\mathrm{O}}$/$L_{\mathrm{IR}}$ of strong-AGN and
 H{\footnotesize{II}}+mild-AGN-dominated sources are shown in red and blue text and dashed lines in the second column.
 $R$ in each panel is the correlation coefficient.
 Mrk231 is shown in red squares.
 Blue and red colors are the same as in Figure \ref{fit}.
 }
 \label{h2o_2_lir}
 \end{figure}

We then analyzed the correlation between $L_\mathrm{H_2O}$/$L_\mathrm{IR}$ 
and the IR colors, along with the $L_\mathrm{IR}$ (Figure \ref{h2o_2_lir} and 
Table \ref{line_ratio}). 
We dismiss lines 1, 7 and 8 here for their insignificant statistics.
Hardly has any correlation been found between $L_\mathrm{H_2O}$/$L_\mathrm{IR}$ 
and $f_{60}$/$f_{100}$ (Table \ref{line_ratio}). 
We find, however, that $L_\mathrm{H_2O}$/$L_\mathrm{IR}$ ratios decrease with the increasing 
$f_{25}$/$f_{60}$, with significant correlation coefficients ($R\sim -0.5$). 
A similar correlation has also been found in lines 7 and 8 though with low statistics.
This correlation may be explained by a smaller contribution to the submillimeter H$_2$O 
line excitation from very warm dust radiation 
(dust temperature $T_\mathrm{d}\sim110\,\mathrm{K}$) than from warm dust
($T_\mathrm{d}\sim50\,\mathrm{K}$). 
We also find that line 6 has the largest $R \sim -0.7$, possibly indicating that this 
transition is more sensitive to T$_d$ than others.
There is no significant correlation between $L_\mathrm{H_2O}$/$L_\mathrm{IR}$
and $L_\mathrm{IR}$ except for line 3 ($R\sim0.4$) as shown 
in the second column of Figure \ref{h2o_2_lir} and in Table \ref{line_ratio}. 
This seems to be consistent with the slightly super-linear relation found for 
$L_\mathrm{H_2O-3}$ with $L_\mathrm{IR}$ (Figure \ref{fit}).
The non-variation of $L_\mathrm{H_2O}$/$L_\mathrm{IR}$ with
$L_\mathrm{IR}$ for most lines confirms the validity of the 
near-linear relations in Figure \ref{fit}.

Again, we here separate the sources into strong-AGN and 
H{\footnotesize{II}}+mild-AGN as in Figure \ref{fit}.
It appears that strong AGNs, on average, have higher $f_{25}$/$f_{60}$ compared with the others. 
This is a well-known property of AGN sources that have more very warm dust than starburst 
sources \citep[e.g.,][]{Younger2009}.
However, both strong-AGN and H{\footnotesize{II}}+mild-AGN species show a 
similar trend for the variation of 
$L_\mathrm{H_2O}$/$L_\mathrm{IR}$ with $f_{25}$/$f_{60}$.
Their different IR colors might cause the average value of
$L_\mathrm{H_2O}$/$L_\mathrm{IR}$ in strong AGNs to be slightly lower, about 40\%, 
than in H{\footnotesize{II}}+mild-AGN sources for all lines (Figure \ref{h2o_2_lir} and Table \ref{line_ratio}),
but the difference is hardly significant.

 \begin{figure}
 \epsscale{1.1}
 \plotone{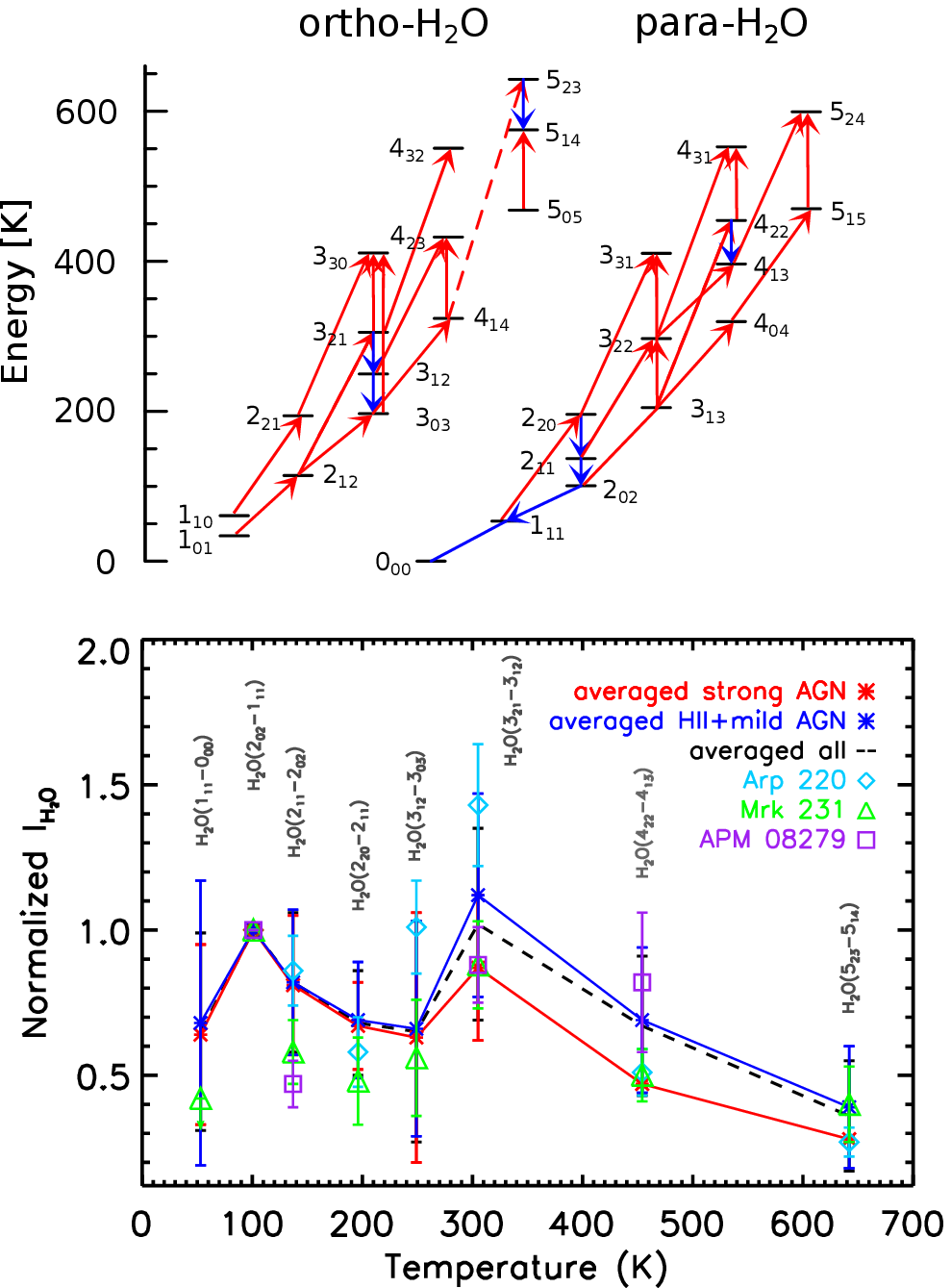}
 \caption{
 Upper panel shows the H$_2$O energy level diagram. 
 Among the red lines that indicate the main possible IR pumping paths, 
 the solid lines show the observed absorption lines in Mrk231, Arp220 and NGC4418 \citep[G-A10;][]{gonz2012}.
 Blue lines are the transitions we detected.
 The lower panel shows the $I_{\mathrm{H}_2\mathrm{O}(2_{02}-1_{11})}$ normalized H$_2$O intensities
 (in $\mathrm{Jy\,km\,s^{-1}}$).
 The black dashed line represents the average values of the whole sample,
 while red and blue points and lines are those of the strong-AGN and 
 H{\footnotesize{II}}+mild-AGN-dominated galaxies, respectively.
 Green and light blue symbols represent Mrk231 (G-A10) and Arp220 \citep{rang2011}, respectively. 
 Purple squares represent the lensed QSO APM08279+5255 \citep{vandw2011}.
 (A color version of this figure is available in the online journal.) }
 \label{sled}
 \end{figure}

\subsection{H$_2$O Line Ratios and the Average Spectral Line Energy Distribution (SLED)}
Line ratios between different transitions could help us understand the excitation of H$_2$O 
and the physical condition of the warm dense gas. 
Thus we compare the H$_2$O line ratios with IR colors and luminosities.
As discussed in Section 2, different transitions have various beam sizes. 
In order to compare different H$_2$O transitions, 
we have to remove this beam size dependence. 
We simply do this by dividing $L_\mathrm{H_2O}$ 
by $L_\mathrm{IR}$ since they are linearly correlated. 
Thus 
$(L_\mathrm{H_2O}$/$L_\mathrm{IR})_{a}$/$(L_\mathrm{H_2O}$/$L_\mathrm{IR})_{b}$ 
(a/b\footnote{a,b represent different H$_2$O lines at different frequencies with different FTS beamsizes.} hereafter)
could represent the true luminosity ratio between two H$_2$O lines, a and b.
Table \ref{line_ratio} lists the results. 
In Figure \ref{h2o_2_lir}, $(L_\mathrm{H_2O}$/$L_\mathrm{IR})_{6}$ 
has the steepest dependence on $f_{25}$/$f_{60}$ compared with other lines.
Thus the ratio between $L_\mathrm{H_2O}$/$L_\mathrm{IR}$ of any other line
and $(L_\mathrm{H_2O}$/$L_\mathrm{IR})_{6}$
should have a correlation with $f_{25}$/$f_{60}$. 
Indeed, as we can see in the table, 
where the $R$ for line ratios 2/6, 3/6 and 5/6 versus $f_{25}$/$f_{60}$ is $\gtrsim$0.5.
Also the line ratio 2/3 decreases with increasing $f_{60}$/$f_{100}$ ($R\sim -0.5$). 
Although there are some $R$ close to $\pm$0.5 for the correlation 
between $\mathrm{I}_\mathrm{H_2O}$/$L_\mathrm{IR}$ 
and $L_\mathrm{IR}$, these trends may not be real for they are within the error. 
The low line ratio 2/6 in Table \ref{line_ratio} might indicate that IR pumping
is important since collisional excitation alone can not explain the 
high intensities of the high-lying lines compared with low-lying lines (e.g., G-A10).

In order to have a general view of the H$_2$O excitation, we calculate 
the error-weighted average line intensity ratios with respect to H$_2$O($2_{02}-1_{11}$). 
In Figure \ref{sled}, the upper panel shows the H$_2$O energy level diagram.
The lower panel of Figure \ref{sled} shows an average H$_2$O SLED together with SLEDs taken from 
previous case studies \citep[G-A10;][]{vandw2011, rang2011}.
The individual studies agree well with our averaged SLED. 
All SLEDs show two peaks at H$_2$O($2_{02}-1_{11}$) and H$_2$O($3_{21}-3_{12}$), 
and the latter is slightly stronger. 
The explanation for the strong high-lying peak could be that the 
IR spectral energy distribution (SED) peaks are close to $75\mu$m which could result in higher IR pumping efficiency
considering the possibility of IR pumping at $75\mu$m (Figure \ref{sled}, upper panel) 
which is the main power source of H$_2$O($3_{21}-3_{12}$) and H$_2$O($3_{12}-3_{03}$) (G-A10).
However, we should be cautious in this interpretation because the 
H$_2$O line intensities depend not only on the excitation conditions, 
but also on the intrinsic line strengths of the H$_2$O molecule.
Detailed excitation modeling is therefore needed.
The high-lying lines to H$_2$O($2_{02}-1_{11}$) ratios
in H{\footnotesize{II}}+mild-AGNs
appear a bit stronger than strong-AGNs (Figure \ref{sled} and Table \ref{line_ratio}).
In Section 3.1, we find that the high-lying lines have steeper anti-correlation on $f_{25}$/$f_{60}$,
thus strong AGNs, with higher $f_{25}$/$f_{60}$, 
are expected to show lower high-lying lines to H$_2$O($2_{02}-1_{11}$) ratios.

\subsection{Emission Lines of H$_2$O-related Ionic and Isotope Molecules}
Besides H$_2$O, the related ionic and $^{18}$O isotope molecular emission lines are also found. 
H$_2$O$^+$ forms via ionization of H and H$_2$, after the combination of H$_2$, it forms
H$_3$O$^+$, and the recombination with electrons leads to OH and H$_2$O \citep{Hollenbach2012}.
Among 45 H$_2$O-detected sources, 5 of them have H$_2$O$^{+}$($1_{11}-0_{00}$,$J_{3/2,1/2}$) (1115.204 GHz), 
another 5 have H$_2$O$^{+}$($1_{11}-0_{00}$,$J_{1/2,1/2}$) (1139.561 GHz),
12 of them have H$_2$O$^{+}$($2_{02}-1_{11}$,$J_{3/2,3/2}$) (746.194 GHz), 
7 sources have H$_2$O$^{+}$($2_{02}-1_{11}$,$J_{5/2,3/2}$) (742.033 GHz) and
3 have H$_2^{18}$O($3_{21}-3_{12}$) (1136.704 GHz) detected. 
Both strong-AGN- and H{\footnotesize{II}}+mild-AGN-dominated galaxies
are among these detections.
We find their luminosities to be tightly correlated with those of the related H$_2$O transitions. 
Taking H$_2$O$^{+}$($2_{02}-1_{11}$,$J_{3/2,3/2}$) that has the largest 
number of detections for an example, the luminosities of H$_2$O$^{+}$($2_{02}-1_{11}$,$J_{3/2,3/2}$) 
and H$_2$O($2_{02}-1_{11}$) perfectly fit a linear correlation. 
H$_2$O$^{+}$($2_{02}-1_{11}$) lines are about 4.5 times weaker than H$_2$O($2_{02}-1_{11}$), 
and 2.5 times weaker than H$_2$O($2_{11}-2_{02}$), while the strength of H$_2$O$^{+}$($1_{11}-0_{00}$) 
is almost the same as that of H$_2$O($1_{11}-0_{00}$). 
These preliminary results are important for further observations of those ionic diagnostic
lines in high-$z$ galaxies, although the number of the sources ($\lesssim10$) is not 
sufficient to draw any concrete conclusion at this stage.

\section{Conclusions}
H$_2$O is found to be the second strongest molecular emitter in our sample 
of 45 nearby IR galaxies after high-$J$ CO lines within the SPIRE/FTS band.
Near-linear correlations have been found between various H$_2$O rotational 
transitions and corresponding $L_\mathrm{IR}$, whereas 
H$_2$O($2_{11}-2_{02}$) and H$_2$O($2_{20}-2_{11}$) may have slightly steeper slopes. 
The ratios of $L_\mathrm{H_2O}$/$L_{\mathrm{IR}}$ vary with $f_{25}$/$f_{60}$, 
while nearly no any trend with $f_{60}$/$f_{100}$ and $L_\mathrm{IR}$ has been found, 
indicating that very warm dust contributes little to the H$_2$O excitation. 
The near constant $L_\mathrm{H_2O}$/$L_{\mathrm{IR}}$ ratios reveal an intrinsic 
linear correlation, no matter whether a strong AGN is present or not. 
We find no significant difference in the correlation between strong-AGN and star-forming-dominated 
galaxies, although strong AGNs might have slightly smaller average 
ratios $L_\mathrm{H_2O}$/$L_{\mathrm{IR}}$.
And in less than one third of both kinds of galaxies, related ionic H$_2$O$^+$ emission lines have 
been detected, while their strength tightly correlates with that of the corresponding H$_2$O lines. 
H$_2^{18}$O isotope line emission is also detected in three sources.
It seems that the IR pumping at $75\mu$m, the IR SED peak, 
is most important in excitation of high-lying H$_2$O lines in these IR galaxies. 
Nevertheless, detailed modeling is needed, e.g. large velocity gradient or XDR models, 
in order to derive some physical parameters of the H$_2$O excitation and to 
provide a quantitative diagnostic tool of the IR radiation 
field and warm dense gas in galaxies other than CO lines.

\acknowledgments
This research is based on the data from HSA and 
is partially supported by the NSF of China (No. 11173059).

{\it Facility: \facility{Herschel}}

\clearpage

\end{CJK*}
\end{document}